# SOME EXACTLY-SOLVABLE QUANTUM PROBLEMS AND THEIR APPLICATIONS TO HETERO- AND NANO-STRUCTURES WITH NONTRIVIAL TOPOLOGY

[1]G. Konstantinou, [2]K. Moulopoulos

*Department of Physics, University of Cyprus, PO Box 20537, 1678 Nicosia, Cyprus*

[1]ph06kg1@ucy.ac.cy, [2]cos@ucy.c.cy

**FINITE-WIDTH INTERFACE**

Analytical calculations based on a Landau Level (LL) picture are reported for a many-electron system moving inside an interface (with a finite-width Quantum Well (QW)) and in the presence of an external perpendicular magnetic field. They lead to a sequence of previously unnoticed singular features in global magnetization and magnetic susceptibility that give rise to nontrivial corrections to the standard de Haas-van Alphen periods. Additional features due to Zeeman splitting are also reported (such as new energy minima that originate from the interplay of QW, Zeeman and LL Physics) that are possibly useful for the design of quantum devices. A corresponding calculation [1] in a Composite Fermion picture [2] leads to new predictions on magnetic response properties of a fully-interacting electron liquid in a finite-width interface.

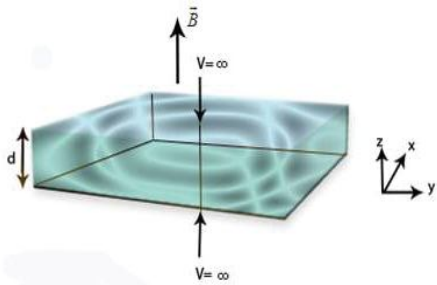

**FIGURE 1:** An interface of finite width **d**, in a perpendicular magnetic field **B**. There is a quantum well in the macroscopic surfaces at z-direction. The energetic competitions between the z-axis levels and Landau Levels (at zero temperature) lead to rapid oscillations of the magnetic properties (energy, magnetization, susceptibility) and force the system to change its character between paramagnetic and diamagnetic; this occurs in a manner that exhibits interesting (and nonintegrable) behavioral patterns with respect to variations of **d** and **B.**

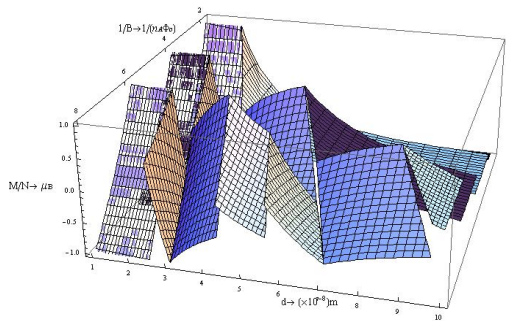

**FIGURE 2:** Magnetization as a function of both **B** and **d**

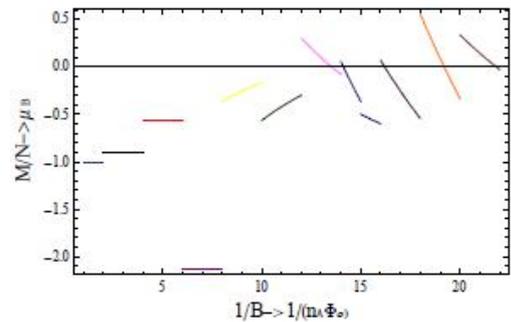

**FIGURE 3:** Magnetization as a function of inverse **B** for fixed width $\left[ d = \sqrt{\dfrac{30\pi}{n_A}} \right]$ ($n_A$ is the areal density)

**FULLY 3-D SYSTEM IN A MAGNETIC FIELD**

We also present exact solutions for the energetics of a fully three-dimensional system of many noninteracting electrons in a magnetic field (a system that has been mostly discussed in astrophysical applications [3]). We find that, at zero temperature, *Hurwitz zeta functions* play an important role on thermodynamic properties (i.e internal energy or magnetization). Although these properties exhibit interesting variations (see Fig. 4 and 5), in the weak magnetic field limit, the de Haas-van Alphen periodicity is rapidly recovered (and so is the well known energy $E = \tfrac{3}{5} N \varepsilon_f$, with $\varepsilon_f$ the 3-D Fermi energy).

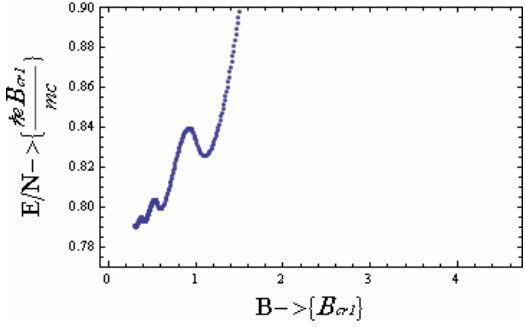 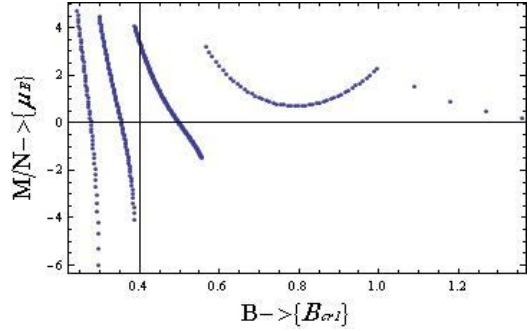

**FIGURE 4:** Energy of 3-D electrons as a function of the magnetic field **B** (determined analytically).

**FIGURE 5:** Magnetization of 3-D electrons as a function of the magnetic field **B** (determined analytically).

In the above figures, the magnetic field is measured in units of $B_{cr1}$, which turns out to be $B_{cr1} = \left(\dfrac{\pi}{16}\right)^{1/3} n^{2/3} \Phi_o$, with $\Phi_o = \dfrac{hc}{e}$ the flux quantum and $n$ the volume density. Magnetization is measured in units of Bohr magneton, $\mu_B = \dfrac{\hbar e}{2mc}$.

## NONTRIVIAL TOPOLOGIES: "ELECTRON GAS-NANOTUBE" AND "1-D INTERFACE"

We finally report on exact solutions for the energetics of an electron gas on a cylindrical surface (and in the presence of an Aharonov-Bohm flux threading the cylinder), where we study two cases: First, we analytically determine the total energy when the radius R of the cylinder is microscopically small (order of nanometers) and we leave the length of the cylinder macrocopically large; second, we study a similar system but now with a microscopic cylinder length (with extra quantum well-confinement) and with a macroscopic radius R. However, all microscopic sizes are formally taken at the end to infinity (in such a way as to maintain constant areal density) and we then recover, in both cases, the well-known 2-D energy $E = \tfrac{1}{2} N \varepsilon_f$, with $\varepsilon_f$ the 2-D Fermi energy. Peculiarities of these two systems are also discussed, especially with respect to emerging Aharonov-Bohm behaviors.

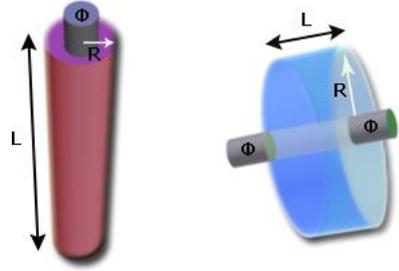

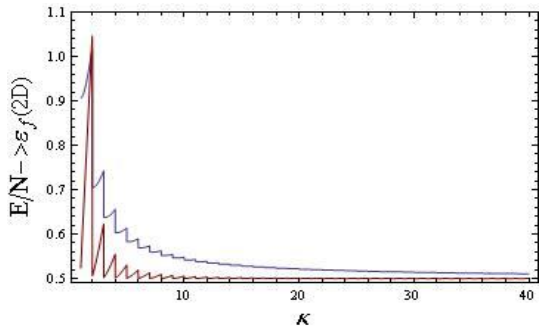

**FIGURE 6:** Energy of electrons (measured in units of 2-D Fermi energy) as a function of the last occupied state in the corresponding microscopic direction (hence parameter $\kappa$ is actually quantized). The red color corresponds to the case when the radius is microscopic and the length is large (nanotube), while the blue color corresponds to the case when the radius is macroscopic and the length is of atomic dimensions (and subjected to a Quantum Well).

Inclusion in the above systems of a radial magnetic field and Zeeman splitting gives rise to curvature-induced spin-orbit coupling [4] that leads to interesting behaviors that can also be dealt with analytically [1]. A corresponding SU(2) formulation and a number of diagonalization results will be presented for the energy of such curved systems that, in the limit of vanishing curvature, have the correct limits (namely, the standard spin-Physics in flat space, that is decoupled from the standard orbital-Physics of Landau Levels).